\RequirePackage{color}
\pdfoutput=1
\documentclass{JINST}
\usepackage{lineno}
\usepackage{graphicx}
\graphicspath{ {./Figures/} }
\usepackage{booktabs}
\usepackage{verbatim}
\usepackage{float} 
\usepackage{amsmath}

\title{Thermal and tensile strength testing of thermally-conductive adhesives and carbon foam}

\author{Maxwell Chertok$^*$, Minmin Fu, Michael Irving, Christian Neher, Mengyao Shi, Kirk Tolfa, Mani Tripathi, Yasmeen Vinson, Ruby Wang, Gayle Zheng\\
University of California, Davis\\
One Shields Ave., Davis, CA 95616, USA\\
\llap{$^*$} 
E-mail: \email{chertok@physics.ucdavis.edu}\\
  }

\abstract{Future collider detectors, including silicon tracking detectors planned for the High Luminosity LHC, will require components and mechanical structures providing unprecedented strength-to-mass ratios, thermal conductivity, and radiation tolerance.  This paper studies carbon foam used in conjunction with thermally conductive epoxy and thermally conductive tape for such applications. Thermal performance and tensile strength measurements of aluminum-carbon foam-adhesive stacks are reported, along with initial radiation damage test results.}

\keywords{Thermal conductivity; Detector mechanics; Silicon tracking detectors}
\begin{document} 

\section{Introduction}
Carbon foam is a material well-suited for particle tracking detector mechanical structure applications due to its relatively high thermal conductivity, structural characteristics such as stiffness and long radiation length, and radiation hardness\cite{snowmass,ibeam,lhcb,plume,allcomp}.
Due to the importance of heat transport in such mechanical structures, bonding methods must maximize thermal conductivity wherever possible.  Thermally conductive epoxies provide good candidates for bonding carbon foam.  However, dedicated characterization of the epoxy--carbon foam interface is not available at present, providing motivation for this work.

The thermal and mechanical properties of adhesives and carbon foam are well known separately, but not when the materials are used in combination.  To create accurate heat dissipation models of components for use in future detectors, experimental data are needed to determine how different combinations of materials affect resulting collective thermal and mechanical properties \cite{snowmass}.  This paper presents measurements of  thermal and mechanical properties of combinations of such materials.  We also report first radiation hardness results for these structures.  

In general, when two or more materials are combined in a single sample such that heat flows sequentially through each material, the thermal properties of the composite are readily calculable.  In the case of the combination of carbon foam and epoxy however, the two components do not remain separated.  Instead, the epoxy is observed to wick via capillary action into the carbon foam to such a degree that this method of prediction is inadequate. Although the thermal conductivities of the carbon foam and epoxy are known separately, it is difficult to predict the thermal conductivity of the combination since heat propagates through the epoxy and carbon foam structure both sequentially and in parallel.  To address this, and to provide input for heat flow simulations, we have fabricated a device to determine the thermal performance experimentally.

Thermally conductive tape is an alternative to epoxy and could potentially be used for bonding elements of detector modules or support structures.  Upon application, thermally conductive tape remains as a solid layer and does not wick into the carbon foam, which results in a weaker bond.  However, this adhesive is desirable for its ease of use and good sample-to-sample repeatability as compared to epoxy and other adhesives.  We present studies of a further alternative adhesive in another paper \cite{rbfpaper}.

Characterizing bond mechanical strength is important to ensure strength and durability of mechanical structures following assembly, especially after substantial radiation exposure.  To this end, we designed a pneumatic tester to measure the ultimate tensile strength of epoxy and tape bonded carbon foam--aluminum stacks.

\section{Apparatus}

The thermal resistance and conductivity of a sample are experimentally determined from the temperature drop $\Delta T$ across the sample when a known amount of power (heat) $P$  flows through it.  Table~\ref{tab:variables} defines these quantities for future reference.
\begin{table}[!h]
\centering
\begin{tabular}{@{}lcc@{}}
\toprule
Quantity             & Symbol & Formula \\ \midrule
Thermal Resistance   & $R$        & $\Delta T/P$       \\
Thermal Conductance  & $C$       & $1/R$       \\
Thermal Conductivity & $k$       & $Cd/A$       \\ \bottomrule
\end{tabular}
\caption{Definitions of resistance, conductance, and conductivity.  $A$ is the cross sectional area of the sample [m$^2$], and $d$ is the thickness of the sample [m].
}
\label{tab:variables}
\end{table}

\subsection{Thermal Conductivity Tester}

\begin{figure}[!ht]
\centering
  \includegraphics[width=0.5\textwidth,natwidth=333,natheight=428]{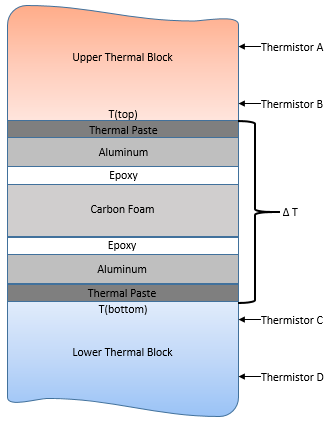}
  \caption{Schematic of stack for thermal testing.  Figure not to scale.}
  \label{fig:stack}
\end{figure}

\begin{figure}[!ht]
\centering
  \includegraphics[width=0.8\textwidth,natwidth=1000,natheight=658]{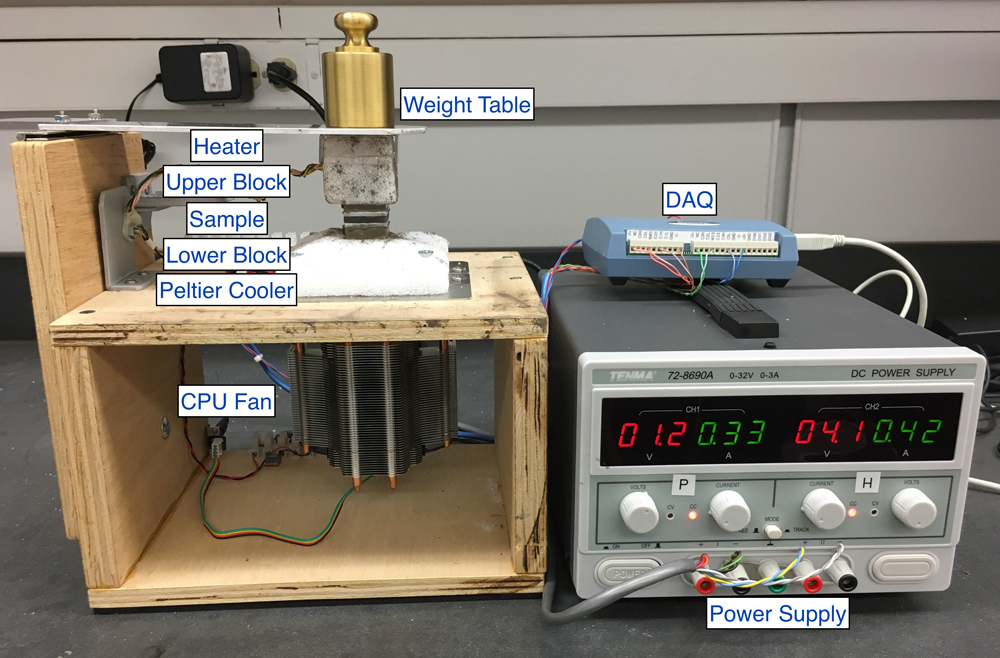}
  \caption{Thermal tester apparatus.  The upper and lower thermal blocks are insulated to reduce heat exchange with the environment.  The DAQ is read out on a PC, not shown.}
  \label{fig:thermaltester}
\end{figure}

For the thermal measurement, we insert the sample (described in the following section) between two instrumented stainless steel blocks to create a vertical stack.  A heater sits on top of the upper block and sends heat through the stack while a Peltier cooler and fan attached to the bottom of the stack draw heat away.  The cooler centers the sample at room temperature in order to minimize heat exchange with the environment.   Figures \ref{fig:stack} and \ref{fig:thermaltester} show a schematic of the stack and photograph of the apparatus.
Four thermistors embedded within the steel blocks track the temperature drop across the sample, with two thermistors in each thermal block. The temperature at each thermistor is calculated using the Steinhart-Hart equation \cite{Steinhart}
\begin{equation}
T = \frac{1}{a + b\ln{(R_e)} + c[\ln{(R_e)}]^3},
\end{equation}
where $R_e$ is the temperature-variable resistance $(\Omega)$ of the thermistor in question and $a$, $b$ and $c$ are the Steinhart-Hart constants included in the manufacturer's specifications for the thermistors.

A 24-bit temperature measurement USB data acquisition device (DAQ) from Measurement Computing collects the temperature data at 0.5 s intervals from the four thermistors. They are referred to, from top to bottom, as thermistors $A, B, C,$ and $D$, respectively. By using the locations of the thermistors and assuming a linear temperature profile through each of the blocks,
we calculate the temperature differential across the sample using the following 3 equations:

\begin{equation}
T_{top} = T_B - \frac{T_A - T_B}{y_A - y_B} (y_B - y_{top})
\end{equation}
\begin{equation}
T_{bottom} = T_C + \frac{T_C - T_D}{y_C - y_D} (y_{bottom} - y_{c})
\end{equation}
\begin{equation}
\Delta T = T_{top} - T_{bottom}
\end{equation}

$T_A$, $T_B$, $T_C$, and $T_D$ are the temperatures (K) measured from each of the four thermistors,  $y_A$, $y_B$, $y_C$, and $y_D$ are the vertical positions of the four thermistors, while $y_{top}$ and $y_{bottom}$ denote the positions of the top thermal block's interface and the bottom thermal block's interface respectively.  The positions are shown in Fig.~\ref{fig:stack}. 

\subsection{Ultimate Tensile Strength Tester}

\begin{figure}[!ht]
\centering
  \includegraphics[width=0.5\textwidth,natwidth=900,natheight=1075]{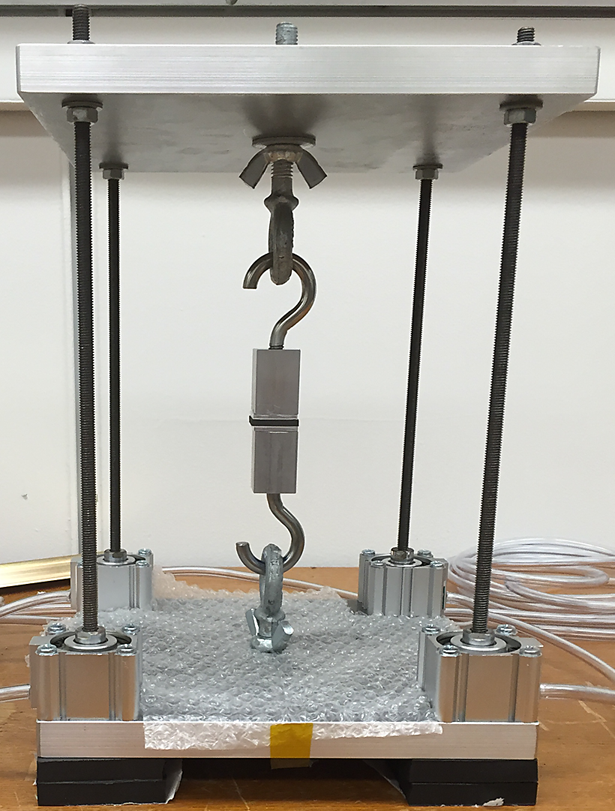}
  \caption{Front view of tensile tester showing pneumatic pistons, frame, and sample mounted with connecting hooks.}
  \label{fig:tensile}
\end{figure}

We have built an apparatus, shown in Figure~\ref{fig:tensile}, to measure ultimate tensile strength (UTS) for these bonded samples.  The tensile tester consists of four pneumatically actuated pistons bolted to the corners of a square base plate.  The pistons push on four rods attached to the corners of an upper plate.  The sample is attached to two aluminum blocks via high strength epoxy.  Hooks are screwed into these blocks, and the hooks are then attached to eye bolts in the center of the upper and lower plates.  The system pressure is gradually increased, and the air flow is stopped at the moment of failure.  At this point, the pressure is recorded using a pressure sensor attached to a manifold.  Calibration is performed by actuating the pistons against known weights, and is well described by a linear function.

\section{Sample Preparation and Testing Procedure}
\begin{figure}[!t]
\centering
  \includegraphics[width=0.4\textwidth,natwidth=790,natheight=822]{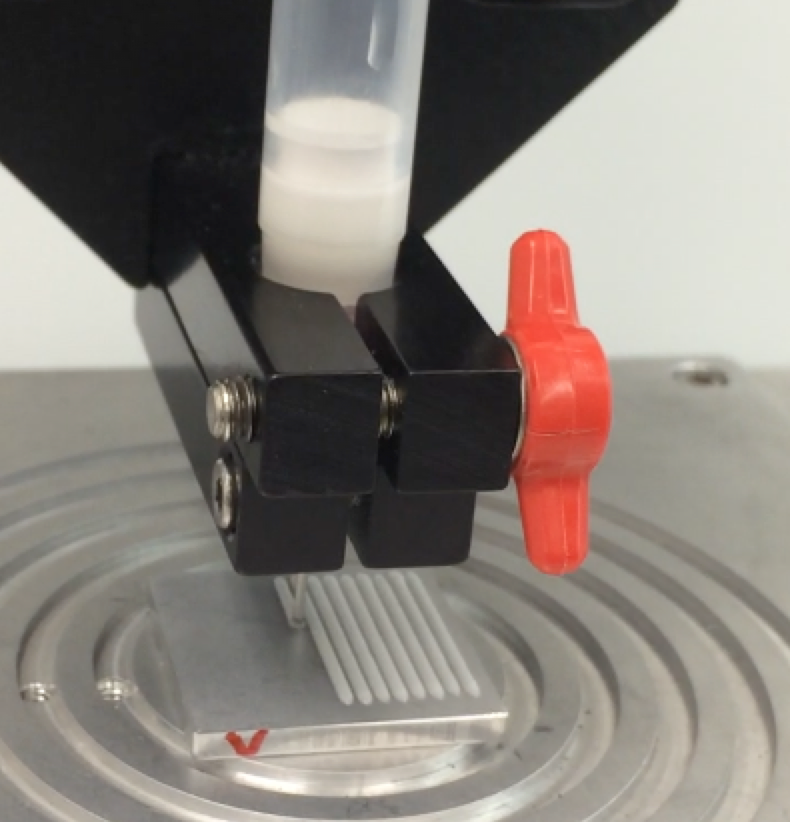}
   \includegraphics[width=0.41\textwidth,natwidth=800,natheight=806]{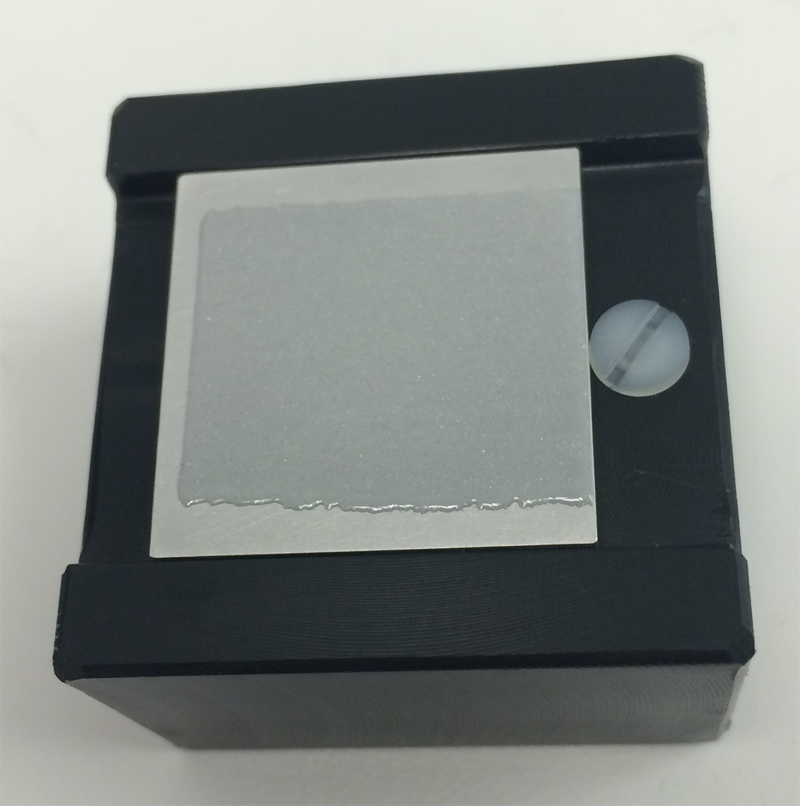}
  \caption{Left: EFD Nordson robotic dispenser applying epoxy to aluminum coupon.  Right: jig and epoxy sample on aluminum coupon after squeegee procedure.}
  \label{nordson}
\end{figure}

Samples for thermal and mechanical testing are stacks with two 25 mm x 25 mm x 2.5 mm 6061 polished aluminum coupons above and below a 25 mm x 25 mm x 4 mm piece of carbon foam, bonded with 3M boron nitride loaded (BN) TC-2810 epoxy or 3M 8800 series conductive tape at each interface between aluminum and carbon foam.  Thermal conductivities (see Table~\ref{tab:variables}) for these materials are: 3M 8805 tape: $k = 0.6$ $W/m \cdot K$ \cite{tape}; 3M BN epoxy TC-2810: $k = 1.0 - 1.4$ $W/m \cdot K$ \cite{epoxy}; 10\% dense carbon foam: $k = 26$ $W/m \cdot K$ \cite{allcomp}.  
An EFD Nordson robotic dispenser applies the epoxy to the aluminum coupon in lines at a speed of 10 mm/s.  (Figure~\ref{nordson}.)  The sample is then placed into a jig of specified depth, and a teflon squeegee is drawn over the sample and edge of the jig.  This leaves a consistent thickness layer of BN epoxy, as shown in  Figure~\ref{nordson}.  After both aluminum coupons are prepared, the carbon foam is inserted in between, a 200 g mass is placed on top of the stack, and the sample is left for 24 hours to cure.  

The thermally conductive tapes used were 3M 8805 and 3M 8810.  The intrinsic properties of these two tapes are identical, but the 3M 8805 tape has a thickness of 125 $\mu$m while the 3M 8810 tape has a thickness of 250 $\mu$m.  We apply the tape to two aluminum coupons of the same dimensions, and form a stack with the carbon foam piece in between.  Following the procedure obtained from 3M, we then apply a pressure of 50 PSI for 5 seconds to ensure an optimal bond.  

\begin{figure}[!ht]
\centering
  \includegraphics[width=0.5\textwidth,natwidth=783,natheight=681]{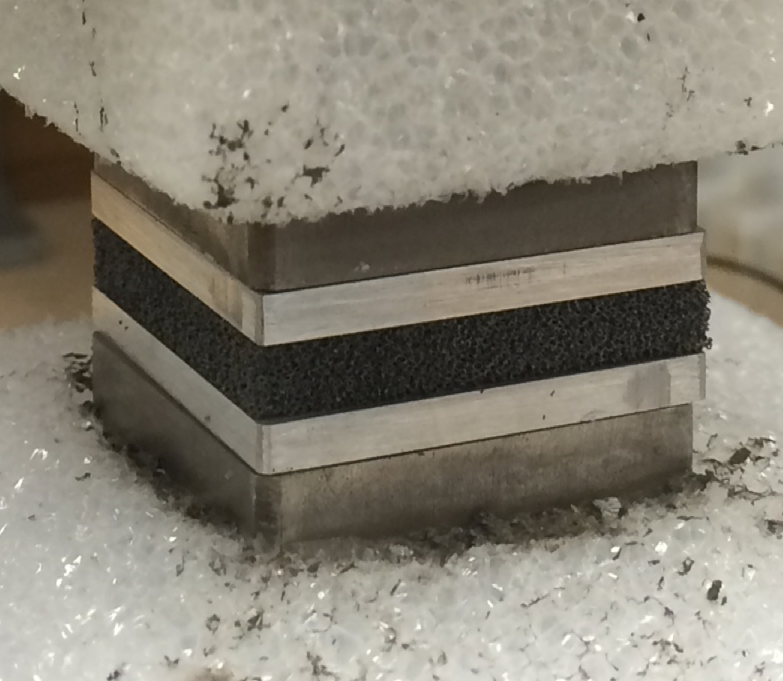}
  \caption{Sample inserted in apparatus, ready for thermal  testing.  The stainless steel blocks with embedded thermistors are encased in insulating foam.}
  \label{fig:samplethermaltest}
\end{figure}

The cured sample is inserted into the thermal tester apparatus, with Arctic Silver 5 thermally conductive paste applied to each sample--stainless steel block interface, as shown in Figure~\ref{fig:samplethermaltest}.  A 1 kg mass is  placed on the weight table (centered directly over the vertical axis passing through the sample stack) and the system is left for 18 hours to settle as the thermally conductive paste spreads out.  Repeated testing of the same sample indicates that variations in the thermal paste's thickness result in no more than 2\% variation in measured thermal resistance.  

	After the requisite settle time, data are collected for 80 minutes, adequate to reach steady state heat flow for this setup.  The heater and cooler are adjusted to ensure the sample is maintained near room temperature and to maximize the temperature difference across the sample, increasing signal to noise.  In practice, 3K is a sufficiently large temperature drop to provide good results without damaging the heater or Peltier.  The data file output by the DAQ is run through a Python script which averages the thermal resistance measured at 70, 75, and 80 minutes.  
	
\section{Results}

\subsection{Thermal Resistance}

Carbon foam samples bonded using BN 3M TC-2810 epoxy or 3M 8800 series tape at a variety of thicknesses were characterized using the thermal conductivity tester described above.  Results are presented in this section.   

\begin{figure}[!ht]
\centering
  \includegraphics[width=0.8\textwidth,natwidth=576,natheight=432]{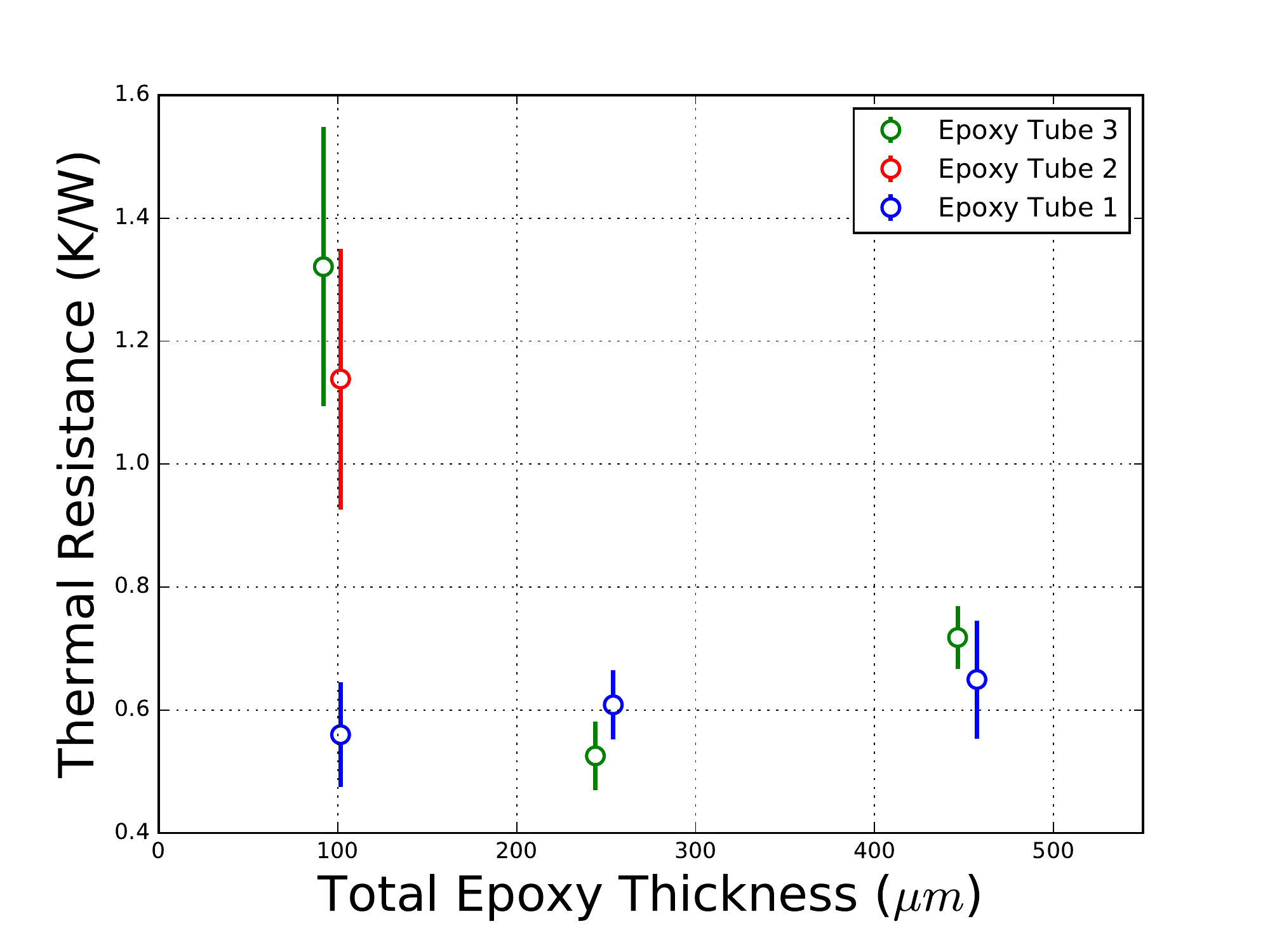}
  \caption{Thermal resistance of BN epoxy--carbon foam samples versus total (2-layer) epoxy thickness.  Results are shown for a variety of samples made with different BN epoxy layer thicknesses and batches.}
  \label{fig:epoxyresults}
\end{figure}

\paragraph{BN Epoxy} 
A series of samples was made using BN epoxy layers of 50 $\mu$m, 125 $\mu$m, and 225 $\mu$m thickness on each interface.  Figure~\ref{fig:epoxyresults} shows the mean thermal resistance of samples  versus total (2-layer) BN epoxy thickness, after subtracting the small contributions from the aluminum and thermal paste, with error bars indicating one standard deviation for the samples made with that thickness.  
Samples were prepared using three different tubes of BN epoxy to study batch-to-batch variations.  Thermal resistance results for samples using the thinnest BN epoxy layers, 50 $\mu$m on each interface, exhibit wide variation, while consistent (i.e., repeatable) values are observed for thicker layers. 

\begin{figure}[!ht]
\centering
  \includegraphics[width=0.7\textwidth,natwidth=1303,natheight=965]{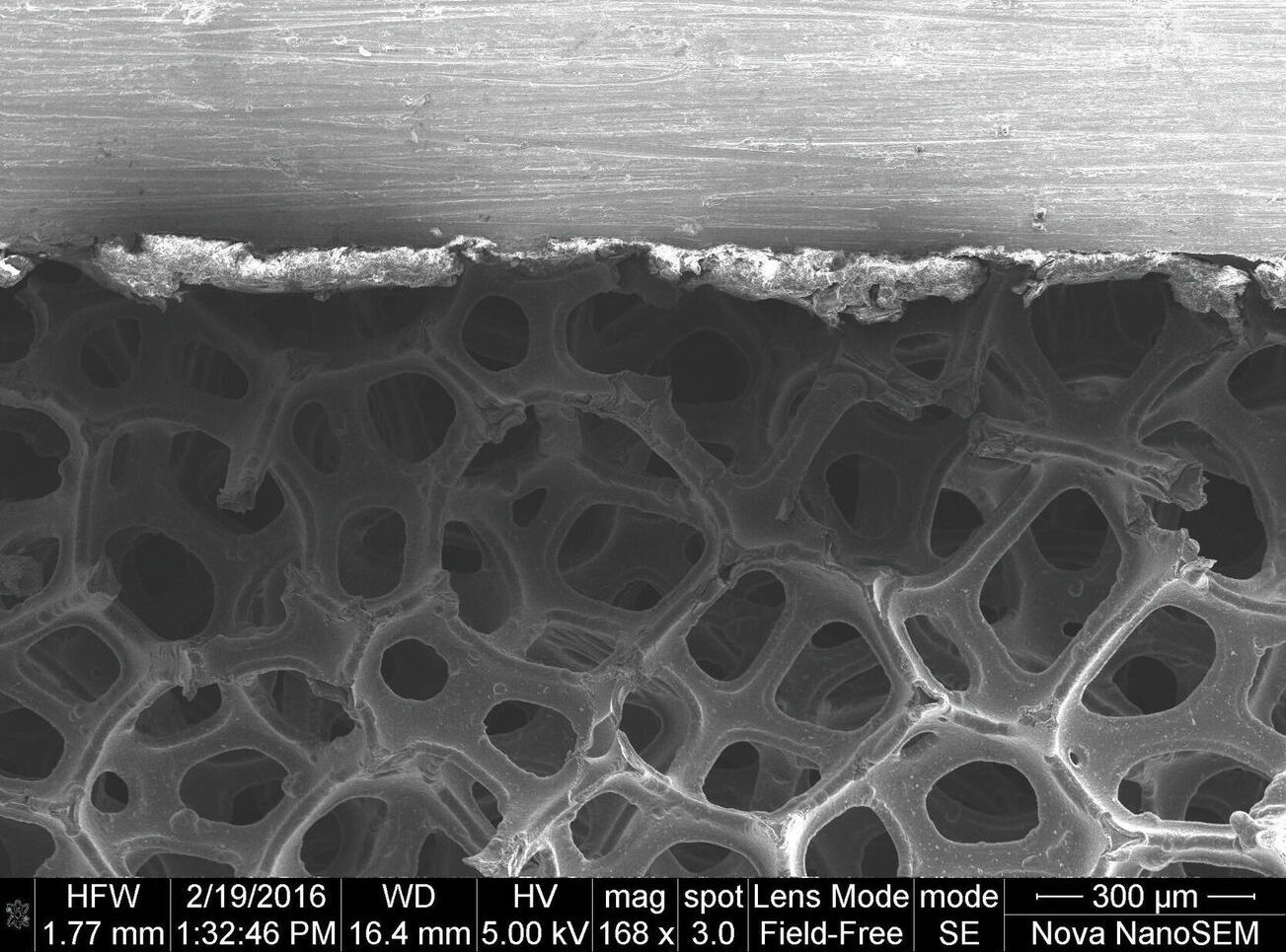}
  \caption{SEM photo of BN epoxy to carbon foam interface.  The BN epoxy is visible as the bright irregular layer between the aluminum and carbon foam.}
  \label{fig:sem}
\end{figure}

 As seen in Figure~\ref{fig:sem}, SEM photos of sectioned samples made with the 50 $\mu$m BN epoxy layer after testing indicate an uneven epoxy layer and inconsistent wicking into the carbon foam. 
As controls, slabs of  BN epoxy without carbon foam were tested, and these showed consistent sample to sample thermal resistance.  Furthermore, tests with only carbon foam and no epoxy also showed consistent results (albeit with high resistance). 

\begin{figure}[!ht]
\centering
  \includegraphics[width=0.7\textwidth,natwidth=576,natheight=432]{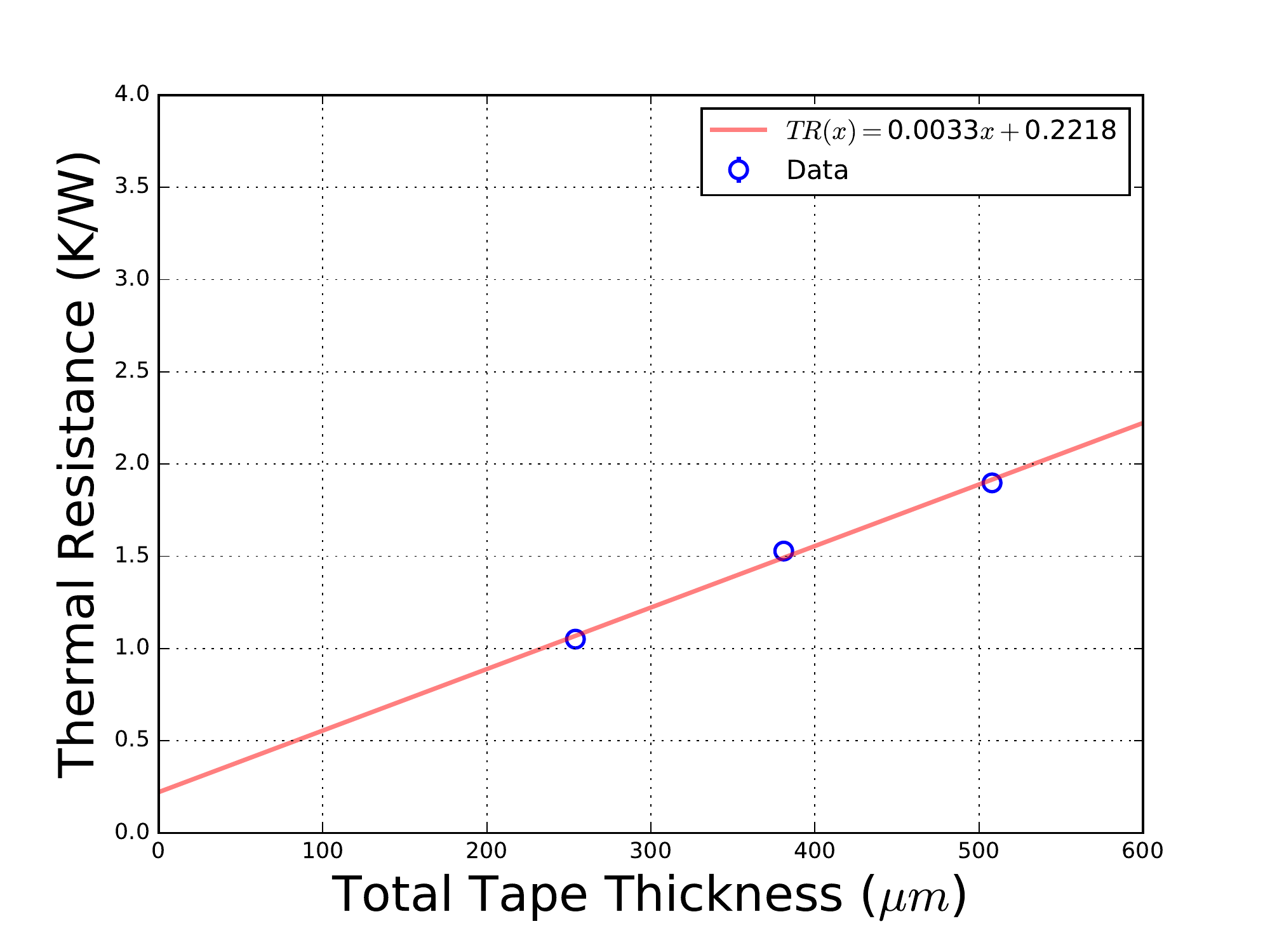}
  \caption{Thermal resistance of tape samples versus total (2-layer) tape thickness.  The red line shows a linear fit to the data.}
  \label{fig:tapegraph}
\end{figure}

\paragraph{Thermally Conductive Tape}
A series of samples was made with $250\mu$m,  $375\mu$m, and $500\mu$m total thickness of 3M 8800 series thermally conductive tape.  Thermal testing shows a linear correlation between sample resistance and tape thickness, as shown in Figure~\ref{fig:tapegraph}.

By fitting a line of the form $TR(x) = ax + b$, we isolate the thermal conductivity of the tape from that of the carbon foam.  This gives
$TR(x) = 0.0033x + 0.2218$, and $k = 0.47$ $W/m \cdot K$ for the thermal tape, 22\% lower than the manufacturer specification of 0.6 $W/m \cdot K$ \cite{tape}.  The second term of the fitted equation gives the thermal conductivity of the carbon foam,   
$k_{carbon} = C_{carbon} \cdot 0.004m / (0.0254m)^2 = 27.96$ $W/m \cdot K$,
in agreement with the specification of 26 $W/m \cdot K$.

\subsection{Tensile Testing}
\begin{figure}[!ht]
\centering
  \includegraphics[width=0.6\textwidth,natwidth=1103,natheight=560]{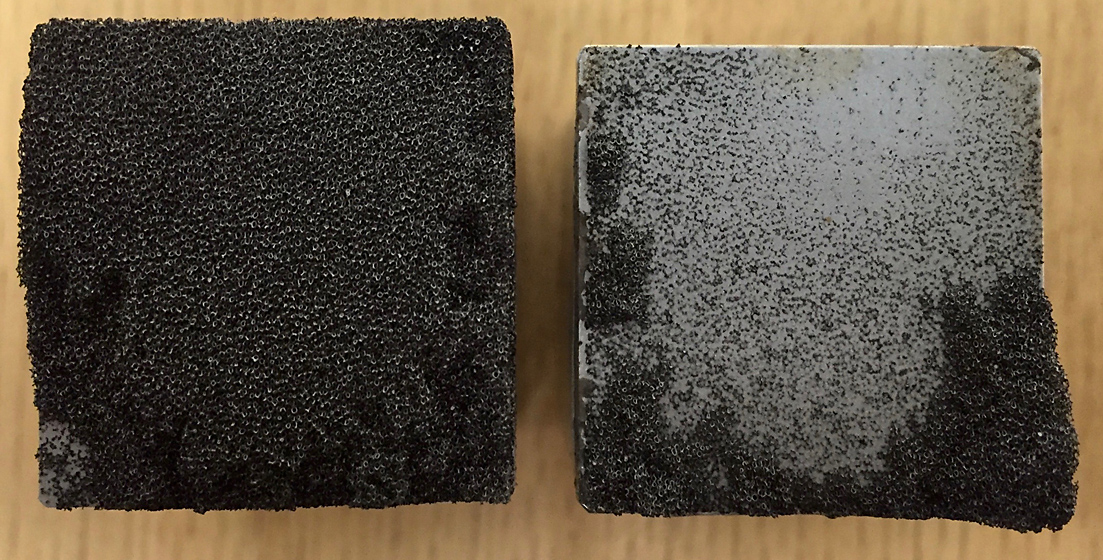}
  \caption{BN epoxy interface after tensile failure.}
  \label{fig:epoxybroken}
\end{figure}

\begin{figure}[!ht]
\centering
  \includegraphics[width=0.6\textwidth,natwidth=1100,natheight=562]{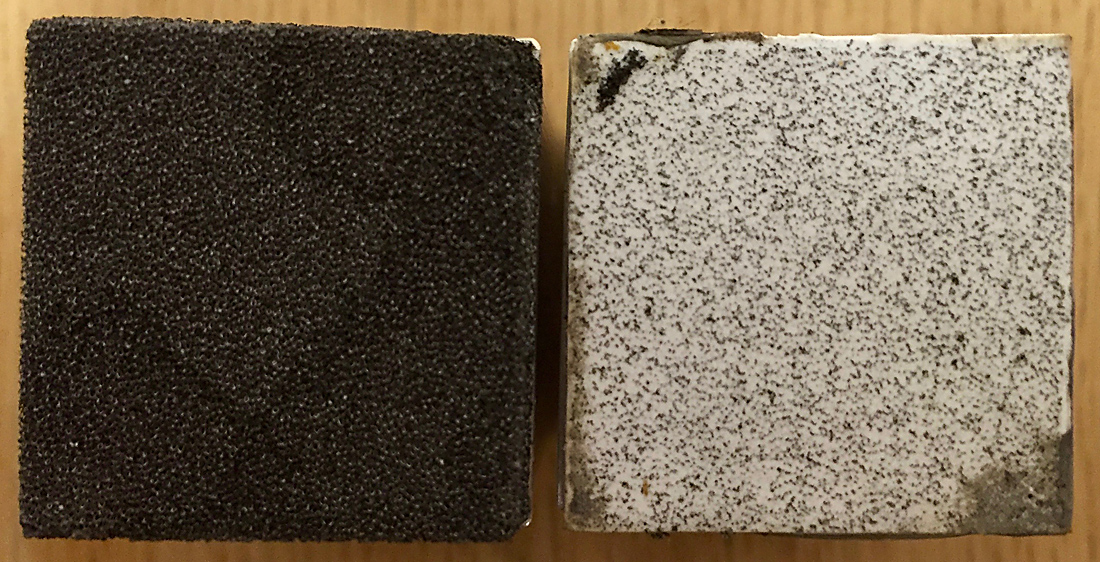}
  \caption{Thermally conductive tape interface after tensile failure.}
  \label{fig:tapebroken}
\end{figure}

Samples of both types were tested for ultimate tensile strength as described above, and the results are shown in Table~\ref{tab:tensile}.   As seen in Figure~\ref{fig:epoxybroken}, the BN epoxy samples after tensile failure contain sizable pieces of carbon foam broken and lifted along with the aluminum coupon.  This indicates that the carbon foam structure breaks before the BN epoxy interface in certain areas.  The 3M 8800 tape tends to stay fully attached to the aluminum coupon, and only lifts traces of the carbon foam surface when broken (Figure~\ref{fig:tapebroken}).  No correlation between UTS and previous heating of the samples (from the thermal testing described above) is observed.

\begin{table}[!ht]
\centering
\begin{tabular}{@{}lllllll@{}}
\toprule
Sample    & Type  & Layer Thickness   & \begin{tabular}[c]{@{}l@{}}Sample 1 \\ (kPa)\end{tabular} & \begin{tabular}[c]{@{}l@{}}Sample 2 \\ (kPa)\end{tabular} & \begin{tabular}[c]{@{}l@{}}Sample 3 \\ (kPa)\end{tabular} & \begin{tabular}[c]{@{}l@{}}Sample 4 \\ (kPa)\end{tabular} \\ \midrule
3M 8805   & Tape  & $125\mu$m   &           219                                                       &             243                                      & 249  & 276\\
3M 8810   & Tape  & $250\mu$m   &                       332                          &                           372                               &  233     & 255\\
3M TC-2810 & Epoxy & $225\mu$m &                  1234                                           &                  742                           &  >1273    &   \\\bottomrule
\end{tabular}
\caption{Ultimate tensile strength (UTS) of tape and BN epoxy samples.}
\label{tab:tensile}
\end{table}

\subsection{Radiation Damage Testing}
One sample, bonded with BN epoxy, was subjected to $10^{14}$ neutrons/cm$^2$ neutron fluence at the McClellan Nuclear Research Center.  The thermal resistance measured before and after irradiation were 1.33 W/K and 1.31 W/K respectively.  This small difference is within the uncertainties of measurement.  UTS for this sample was determined as described above, and results are consistent with those in Table~\ref{tab:tensile}.


\section{Discussion}

\paragraph{Thermal resistance measurements}
These results indicate that while 3M TC-2810 BN epoxy has relatively high thermal conductivity, its thermal properties when used to bond carbon foam can be  inconsistent from sample to sample, even for samples made with identical conditions.  This is apparent for the thinnest 50 $\mu m$ BN epoxy layers.  It is likely that differences in secondary properties such as temperature, humidity, and viscosity influence wicking into the carbon foam, and thus greatly affect the thermal properties of the samples at this thickness.   We note that the data sheet provided by 3M indicates a $\sim~40\%$ batch-to-batch range in thermal conductivity \cite{epoxy}.  

Repeatable results were obtained for BN epoxy layers of  125 $\mu m$ and above.  Thus, there is a trade off between consistency and higher thermal conductivity as the BN epoxy is less conductive than the carbon foam under study.  
Samples made with 225 $\mu$m BN epoxy layers exhibit slightly worse thermal performance than those made with 125 $\mu m$ layers.  This is as expected, due to the increased thermal resistance of the thicker layer.  However, these 225 $\mu$m samples were still $\sim$ 15\% more conductive than would be expected for unmixed layers of BN epoxy and carbon foam in series.  Thus, wicking is seen to improve the thermal performance for layers of this thickness.  

Samples made with 3M 8800 series thermally conductive tape showed consistent sample-to-sample results with a highly linear relationship between tape thickness and thermal resistance.  However, the thermal performance of these samples is considerably worse than those made with the BN epoxy.  

\paragraph{UTS measurements}
The BN epoxy samples showed an up-to 66\% difference in UTS between samples made using the same well-controlled procedure.  Despite this inconsistency, even the weakest of the BN epoxy samples was still greater than twice as strong as any tape sample tested.  
Failure mode analysis indicates this is partially due to the fact that the tape cannot wick into the foam structure like the epoxy does.  A modest increase in UTS is observed by doubling the thickness of the thermally conductive tape at the expense of nearly doubling thermal resistance.

\paragraph{Radiation Damage measurements}
Radiation damage testing of BN epoxy--foam samples performed with neutrons to a fluence of $10^{14}$ n/cm$^2$ showed no change in thermal or mechanical properties.  Tests with charged particles and at higher fluences are necessary to ensure suitability  at harsh radiation environments, such as at High Luminosity-LHC.

\section{Conclusions and Future Work}
We have investigated the use of thermally conductive BN epoxy and tape to bond carbon foam  for applications such as those envisaged for particle physics tracking detector mechanical structures.  BN epoxy provides better thermal conductivity and ultimate tensile strength, although thin layers exhibit substantial inconsistencies of response.  Thermally conductive tape can be a viable alternative depending on the application.  Initial  radiation damage testing with neutrons shows the BN epoxy--carbon foam interface is robust, although more testing is required.  We plan further studies with  complex configurations, such as for systems with cooling pipes embedded in the carbon foam.

\section{Acknowledgments}
This work at the University of California, Davis was supported by U.S. Department of Energy grant DE-FG02-91ER40674 and by U.S. CMS R\&D funds via Fermilab.


\begin{thebibliography}{9}
\bibitem{snowmass} M. Artuso, et al, "Sensor Compendium - A Snowmass Whitepaper,"  http://arxiv.org/abs/1310.5158, [physics.ins-det], 2013.

\bibitem{ibeam} N. Hartman, et al, "Novel fabrication techniques for low-mass composite structures in silicon particle detectors," Nuclear Instruments and Methods in Physics Research A 732 (2013) 103 - 108.

\bibitem{lhcb} LHCb Collaboration, "LHCb Tracker Upgrade Technical Design Report," CERN-LHCC-2014-001 ; LHCB-TDR-015, 2014.

\bibitem{plume} PLUME Collaboration, "Ultra-light ladders for linear collider vertex detector," Nucl. Inst. Meth. A, Vol 650, 1, 2011.


\bibitem{allcomp}
Allcomp corp., http://www.allcomp.net/.

\bibitem{tape}
3M Electronics Materials Solutions Division,
	"3M Thermally Conductive Adhesive Transfer Tapes 8805, 8810, 8815, 8820", 2015.

\bibitem{epoxy}
3M Electronics Materials Solutions Division,
"3M Thermally Conductive Epoxy Adhesive TC-2810", 2014.

\bibitem{rbfpaper}  M. Chertok, et al,  "Reactive Bonding Film for Bonding Carbon Foam Through Metal Extrusion," http://arxiv.org/abs/1606.07677, [physics.ins-det], 2016.  Submitted to JINST.

\bibitem{Steinhart} J. S. Steinhart and S. R. Hart,
"Calibration curves for thermistors", Deep Sea Research and Oceanographic Abstracts, volume 15, number 4, 1968.



\end{thebibliography}
\bibliographystyle{unsrt}

\end{document}